\documentclass[preprint2]{aastex}
\usepackage{aastexug}
\begin{document}

\title{Giant Molecular Outflows Powered by Protostars in L1448}

\author{\sc 
Grace A. Wolf-Chase\altaffilmark{1,2},
Mary Barsony\altaffilmark{3,4,},
and JoAnn O'Linger\altaffilmark{5} 
}

\altaffiltext{1}{Dept. of Astronomy \& Astrophysics, University of Chicago,
5640 S. Ellis Ave., Chicago, IL 60637; grace@horta.uchicago.edu}
\altaffiltext{2}{Adler Planetarium \& Astronomy Museum,
1300 S. Lake Shore Drive, Chicago, IL 60605}
\altaffiltext{3}{NSF POWRE Visiting Professor, Physics Department, Harvey Mudd
College, Claremont, CA  91711; barsony@home.com}
\altaffiltext{4}{NSF CAREER Award Recipient}
\altaffiltext{5}{Jet Propulsion Laboratory, 4800 Oak Grove Dr., MS 100-22,
Pasadena, CA 91109; joanno@ipac.caltech.edu}

\begin{abstract}

We present sensitive (T$_R^*\ \approx\ $0.1K), 
large-scale (47$^{\prime}$ $\times$ 7$^{\prime}$--corresponding to 4 pc $\times$
0.6 pc
at the source) maps of the  CO J=1$\rightarrow$0 emission
of the L1448 dark cloud at 55$^{\prime\prime}$ resolution.
These maps were acquired using the On-The-Fly (OTF) capability of the
NRAO
12-meter telescope atop Kitt Peak in Arizona.
CO outflow activity is seen in L1448 on parsec-scales for the first time.
Careful comparison of the spatial and velocity
distribution of our high-velocity CO maps
with previously published optical and near-infrared images and spectra has 
led to the identification of six distinct
CO outflows. Three of these are powered by the Class 0 protostars,
L1448C, L1448N(A), and L1448N(B).
L1448 IRS 2 is the source of two more outflows, one of which
is newly identified from our maps.
The sixth newly discovered outflow is powered by an as yet unidentified source
outside of our map boundaries.

We show the direct link between the heretofore unknown, giant,
highly-collimated,
protostellar molecular outflows and
their previously discovered, distant optical manifestations.
The outflows traced
by our CO mapping generally reach 
the projected cloud boundaries.
Integrated intensity maps over narrow velocity intervals indicate there
is significant overlap of blue- and redshifted gas, suggesting the outflows
are highly inclined with respect to the line-of-sight, although the individual
outflow position angles are significantly different.
The velocity channel maps also show
that the outflows dominate the CO line cores as well as the high-velocity wings.
The magnitude of the combined flow momenta,
as well as the combined kinetic energy of the
flows, are sufficient to disperse the 50 M$_{\odot}$ NH$_3$ cores in which
the protostars are currently forming, although some question remains as to
the exact processes involved in redirecting
the directionality of the outflow momenta
to effect the complete dispersal of the parent cloud.

\end{abstract}


\section{Introduction}

It has long been an open question whether young stars could be the agents of
dispersal of their parent molecular clouds through the combined effects
of their outflows \citep{nor79,ber96}.
The answer to this question depends on whether the outflows have the 
requisite kinetic energy to overcome the gravitational binding energy of the
cloud, as well as the efficiency with which outflows can transfer momentum,
in both magnitude and direction, to the surrounding cloud.

For considerations of molecular cloud dispersal,
addressing the question of the adequacy of outflow momenta
has historically lagged behind determinations of outflow energetics.
This is because evaluation of the available
energy sources needed to account for the observed spectral linewidths in a cloud
is adequate for quantitative estimates of outflow energies. However,
in order to address whether the requisite momentum for cloud dispersal exists in
a given case requires well-sampled, sensitive, large-scale mapping of
sufficiently
large areas to encompass entire molecular clouds. Such observing
capability has been beyond reach until the last few years, with the 
implementation of  ``rapid'' or ``On-The-Fly'' mapping capabilities at 
large-aperture millimeter telescopes.

The fact that many outflows powered by young stellar objects actually
extend well beyond their parent molecular cloud boundaries has been
recognized only recently, with the advent of large-scale, narrowband optical
imaging
surveys that have revealed shock-excited Herbig-Haro objects 
at parsec-scale separations from their exciting sources
\citep{ba96a,ba96b,bal97,dev97,eis97,wil97,gom97,gom98,rei98}
and from equally large-area, sensitive, millimeter line maps that show
parsec-scale
molecular outflows \citep{den95,lad96,ben96,ben98,oli99}.
The millimeter line maps of parsec-scale flows have been almost 
exclusively confined to instances
of single, well-isolated cases, due to the tremendous confusion of multiple
outflows in regions of clustered star formation, such as are found in 
NGC 1333 (Sandell \& Knee 1998; Knee \& Sandell 2000),
$\rho$ Oph, Serpens \citep{whi95}, or Circinus \citep{bal99}.

The L1448 dark cloud, with a mass 
of 100 M$_{\odot}$ over its $\sim$ 1.3 pc $\times$ 0.7 pc extent as traced
by C$^{18}$O emission, \citep{ba86a}, is part of the much more extensive
(10 pc $\times$ 33 pc) Perseus molecular cloud complex, which contains
$\approx$ 1.7 $\times$ 10$^4$ M$_{\odot}$, at a distance of
300 pc \citep{ba86b}. The two dense ammonia cores within L1448
contain 50 M$_{\odot}$ distributed over a 1 pc $\times$ 0.5 pc
area \citep{ba86a,ang89}. The core at V$_{LSR}$ = 4.2 km s$^{-1}$ 
contains the Class 0 protostar L1448 IRS 2, while
the other core, at V$_{LSR}$ = 4.7 km s$^{-1}$, harbors four Class 0
protostars: L1448C, L1448N(A), L1448N(B), and L1448NW.
\citep{bar98,oli99,eis00}. The Class I source,
L1448 IRS 1, lies close to the western boundary of the cloud, just outside
the lowest NH$_3$ contours in the maps of \citep{ba86b}.

High-velocity molecular gas in L1448 was discovered a decade ago
via CO J$=$2$\rightarrow$1 and CO J$=$1$\rightarrow$0 mapping of a 
$\sim$ 2$^{\prime}$ $\times$ 6$^{\prime}$ area centered on L1448C, 
acquired with 12$^{\prime\prime}$ and 20$^{\prime\prime}$ angular resolutions,
respectively \citep{bac90}.
Due to its brightness, high-velocity extent ($\pm$ 70 km s$^{-1}$), and 
symmetrically spaced CO bullets, the L1448C molecular outflow has been the
object of much study, unlike the flows from its neighbors, the
7$^{\prime\prime}$
(in projected separation) protobinary, L1448N(A) \& (B),
just 1.2$^{\prime}$ to the north, or L1448 IRS 2, 3.7$^{\prime}$ to the
northwest (e.g., \citep{cur90,gui92,bal93,bac94,dav94,bac95,dut97}).
Although outflow activity in the vicinity of the protobinary
had been reported  previously,
the H$_2$ and CO flows, driven by L1448N(A) and L1448N(B),
respectively \citep{bac90,dav95},
were not recognized as distinct until recently \citep{bar98}.
Identification of these flows was aided by noting the position angle
of the low-excitation H$_2$ flow, centered on L1448N(A), 
to be distinct from the position angle of the CO flow from L1448N(B),
defined by the direction of the line joining L1448N(B) with
the newly discovered Herbig-Haro object, HH 196 \citep{bal97}.

Recent, wide-angle ($\sim$ 70$^{\prime}$ field-of-view), narrowband optical
imaging of the entire extent of the L1448 cloud has resulted in the
discovery of several systems of Herbig-Haro objects, some displaced
several parsecs from any exciting source \citep{bal97}.
In order to investigate the link between high-velocity molecular gas and
the newly discovered Herbig-Haro objects, as well as to study
the possibility of cloud dispersal via outflows,
we acquired new, sensitive, large-scale CO J$=$1$\rightarrow$0
maps of a substantial portion of the L1448 cloud. These new molecular line
maps were acquired with the On-The-Fly (OTF) mapping 
technique as implemented at NRAO's 12-meter millimeter telescope atop
Kitt Peak, Arizona.
 
\section
{Observations and Data Reduction}

The CO J$=$1$\rightarrow$0 maps of L1448 presented in this paper
were acquired using the spectral-line On-The-Fly (OTF) mapping mode
of the
NRAO's\footnote{The National Radio Astronomy Observatory is a facility of
the National Science Foundation, operated under cooperative agreement by
Associated
Universities, Inc.} 12-meter telescope on 23 June 1997,
UT 13$^h$53$^m$ $-$ UT 19$^h$25$^m$.
We stress that the OTF technique allows the acquisition of large-area,
high-sensitivity, spectral line maps with unprecedented speed and pointing 
accuracy.  For comparison, it would have taken eight
times the amount of telescope time, or nearly a week in practice,
to acquire this same map using conventional, point-by-point mapping.
Although OTF mapping is not a new concept, 
given the rigor of the position encoding that allows precise and accurate
gridding of the data, the fast data recording rates that allow rapid scanning 
without beam smearing, and the analysis tools that are available,
the 12-meter implementation is the most ambitious effort at OTF imaging yet.

To produce our CO maps of L1448, we observed a 47$^{\prime}$
$\times$ 7$^{\prime}$ field along a position angle P.A. $=$ 135$^{\circ}$,
(measured East from North), centered on
the coordinates of L1448 IRS 2 ($\alpha_{1950}=$ 3$^h$ 22$^m$ 17.9$^s$, 
$\delta_{1950}=$ 30$^{\circ}$ 34$^{\prime}$ 41$^{\prime\prime}$).
The 115 GHz beamwidth was $\approx$ 55$^{\prime\prime}$.
We scanned a total of 33 rows
at a rate of 50$^{\prime\prime}$/s, along
P.A. $=$ 135$^{\circ}$,
with a row spacing of 12.7$^{\prime\prime}$.
(The row spacing is determined by the optimum spatial sampling and by the
scanning position angle.)
We calibrated and integrated
on an absolute off position
($\alpha_{1950}=$ 3$^h$ 20$^m$ 00.0$^s$, 
$\delta_{1950}=$ 31$^{\circ}$ 00$^{\prime}$ 00$^{\prime\prime}$)
at the start of every row. Each row took approximately one minute to scan.
Each map coverage took 41 $-$ 56 minutes to complete.
We performed six map coverages to attain an RMS in each OTF spectrum of
T$_R^*$ $\approx$  0.11 K.

A dual-channel, single-sideband SIS receiver was used for all observations.
The backend consisted of 
250 kHz and 500 kHz resolution filterbanks, yielding velocity resolutions of
0.65 km s$^{-1}$ and 1.3 km s$^{-1}$, respectively. The filterbanks 
were used in parallel mode, each of the two receiver polarization channels
using 256 filterbank channels.
The polarization channels were subsequently averaged
together to improve signal-to-noise.
Only the 250 kHz resolution data were used to produce the maps presented here.

Line temperatures at the 12-meter are on the T$_R^*$ scale, and must be divided
by
the
corrected main beam efficiency, $\eta_m^*$, to convert to the main-beam
brightness 
temperature scale.  For our very extended source, $\eta_m^*$ is approximately
1.0.
Since the corrected main-beam efficiency is the fraction of the forward power
in the main diffraction beam relative to the total forward power in the 
main beam plus error beam, contribution from the error beam can make 
$\eta_m^*$ $>$ 1.0.  At 115 GHz, the theoretical error beam width is 
$\approx$ 17$^{\prime}$, but the ratio of the error beam amplitude to the main
beam
amplitude is only 6 $\times$ 10$^{-4}$, suggesting contribution from the 
error beam can be ignored.   
We used the NRAO standard source, B5 
($\alpha_{1950}=$ 3$^h$ 44$^m$ 52.4$^s$, 
$\delta_{1950}=$ 32$^{\circ}$ 44$^{\prime}$ 28$^{\prime\prime}$),
to check absolute line temperatures.

The OTF data were reduced with the Astronomical Image Processing
Software (AIPS), Version 15JUL95.
AIPS tasks specific to OTF data 
are `OTFUV', which converts a single 12-meter OTF map (in UniPOPS SDD
format) to UV (single-dish) format,
and `SDGRD', which selects random position single-dish data in AIPS UV format
in a specified field of view about a specified position and projects the
coordinates onto the specified coordinate system.
The data are then convolved onto a grid.
OTF data maps were first combined, then gridded into a
data cube and baseline-subtracted.
Channel maps as well as
individual spectra were inspected
to ensure good baseline removal and to check for scanning artifacts. Only the 
first and last rows scanned contained corrupted spectra
and were rejected.

\section{Results}

\subsection{The Extent of High-Velocity CO in L1448}

Figure 1a shows the extent of high-velocity blue- and redshifted
CO J$=$1$\rightarrow$0 emission found within
our OTF map boundaries, outlined by the zig-zag lines.
The red contours represent high-velocity redshifted emission integrated over
the velocity interval 8.1 km s$^{-1}$ $\le$ V$_{LSR}$ $\le$ 17.8 km s$^{-1}$,
whereas the blue contours represent high-velocity blueshifted emission
integrated over
$-$12.1 km s$^{-1}$ $\le$ V$_{LSR}$ $\le$ $-$1 km s$^{-1}$.
The rest velocities of the two NH$_3$ cores are 4.2 km s$^{-1}$ and 4.7
km s$^{-1}$.
The mapped area comprises 47$^{\prime}$ $\times$ 7$^{\prime}$, 
corresponding to 4 pc $\times$ 0.6 pc at the source, at a distance of 300 pc.
Stars indicate the positions of the known Class 0 sources in L1448,
which include L1448C, L1448N (the unresolved, 7$^{\prime\prime}$ separation
protobinary L1448N(A) \& L1448N(B)), L1448NW (20$^{\prime\prime}$ northwest of
the
protobinary), and L1448 IRS 2, at map center. A filled square 
indicates the position of the Class I Young Stellar Object (YSO), L1448 IRS 1
(Cohen 1980;
Eisl\"offel 2000). Figure 1b, the inset to Figure 1a,
shows the scale of the previously mapped region of high-velocity CO,
for comparison (from \citep{bac90}). This earlier map
includes only a small portion of the high-velocity gas associated with Class 0
protostars in L1448.
Comparison of Figures 1a and 1b highlights
the truly spectacular spatial extent of outflow activity in L1448.

Figure 2 indicates the positions of all the known Herbig-Haro objects (crosses)
which are found within our map boundaries, superimposed on the CO map of Figure
1a.
Several striking features are evident in Figure 2: (1) There is 
a $\sim$15$^{\prime}$ long blueshifted filament that is connected to both
the L1448N/L1448C region and to L1448 IRS 2 in a wishbone-shaped structure
which contains HH 197, HH 195A$-$D, HH 196, and 
culminates in HH 193;
(2) There is extended redshifted outflow emission which
suggests structure along three separate axes (at P.A.'s$\sim$129$^{\circ}$,
150$^{\circ}$, \& 180$^{\circ}$) directly to the southeast of
L1448N/L1448C; (3) In the immediate vicinity
of IRS 2, the blueshifted gas peaks on HH 195E, whereas the center of the red
peak is 
along a line drawn through IRS 2 and HH 195A$-$D;
(4) There is redshifted emission that peaks $\sim$9$^{\prime}$ to the
southeast of IRS 2, which lies on a line connecting IRS 2 and HH 193 at
P.A.$\sim$152$^{\circ}$;
(5) There is redshifted emission that peaks on HH 277, which appears to be 
oriented nearly perpendicular to the long axis of our map;
(6) There is blueshifted
CO emission associated with the HH 267 knots, which lie at the northwestern
edge of our map. 

In addition to the OTF map, we also
acquired single-point spectra of the CO J$=$1$\rightarrow$0
and $^{13}$CO J$=$1$\rightarrow$0 transitions 
at four positions, as indicated in Figure 3.

These spectra were used to help determine the appropriate velocity
integration limits for the high-velocity CO emission shown in Figures 1a and 2.
The mapped linewing emission velocities are indicated by the horizontal arrows
for the the blue-shifted emission in Figures 3a \& 3c, and for the 
red-shifted emission in Figures 3b \& 3d.
These spectra were also used
to determine velocity-intervals free of
line emission for baseline-subtraction of the gridded
OTF data, and to check CO optical depths in the line wings. 
The spectra shown in Figure 3a were obtained just off the northwest corner
of our map, near the HH 267 knots, whereas the spectra of
Figures 3b, c, \& d were obtained at positions of strong outflow emission
within our map boundary.
The spectra in Figure 3a
show a separate velocity feature
at 0 km s$^{-1}$ in both CO and $^{13}$CO.
The integration limits for the high-velocity gas were chosen conservatively,
avoiding emission from
the velocity feature at V$_{LSR}=$0 km s$^{-1}$.
It is interesting to note that this velocity 
feature corresponds to the V$_{LSR}$ of three HH objects in this cloud:
HH 196A, HH 196B, \& HH 267B.

\subsection{Background: Individual Outflow Sources}

\subsubsection{The L1448C/L1448N Core}

The CO outflow from L1448C has been studied extensively since its
discovery, at which time it was recognized as a 
unique source due to its high collimation factor
(approaching 10:1) and its extremely high velocities
($\pm$70 km s$^{-1}$ $-$ \citep{bac90,bac95}.
Interferometric observations of the outflow, acquired with a 
3$^{\prime\prime}$ $\times$ 2.5$^{\prime\prime}$ synthesized beam,
were required to resolve the limb-brightened CO cavities at ``low'' velocities
($-$12 $\le$ V$_{LSR}$ $\le$ $+$16 km s$^{-1}$) \citep{bac95,dut97}.
By modelling the interferometric CO channel maps,
the outflow inclination angle was found to be $i\ =\ 70^{\circ}$, 
implying actual jet velocities in excess of 200 km s$^{-1}$.
The initial conical outflow opening half-angle was found to be
$\phi$/2 $=$ 22.5$^{\circ}$ \citep{bac95}.
The outflow cavity walls become parallel, however,
$\approx$ 1$^{\prime}$ ($=$ 0.08 pc) downstream from the driving source,
with a width of $\sim$ 20$^{\prime\prime}$.
Therefore, in our OTF map, the L1448C outflow 
remains unresolved along its width.

The redshifted lobe of the L1448C outflow
is deflected by $\sim$ 20$^{\circ}$ from its initial direction, from an
initial position angle of 
$+$160$^{\circ}$ to a final position angle of $+$180$^{\circ}$ 
\citep{dut97,eis00}.
This change in direction
of the redshifted flow axis occurs abruptly near the position
of the CO ``bullet'' known as R3
\citep{bac90,dut97}. 
A string of H$_2$ emission knots along P.A. $= 180^{\circ}$ extends for nearly
two arcminutes, beginning about an arcminute southeast of L1448C
(Eisl\"offel 2000).
Similarly, the blueshifted CO outflow lobe powered by L1448C
starts out at a position angle
P.A. $=$
$+$159$^{\circ}$ 
\citep{bac95}, before being
deflected through a total angle of $\sim$ 32$^{\circ}$ 
by the time it arrives an arcminute downstream (Davis \& Smith 1995;
Eisl\"offel 2000).  This deflection is due to the collision of the
blueshifted gas driven by L1448C with the ammonia core containing
the protobinary L1448N(A) \& (B) \citep{cur99}.
The strongly radiative shock emission at this interaction
region is evident through various tracers:
enhanced CO millimeter line emission at the site of the CO bullet ``B3'' of
Bachiller et al. (1990),
a far-infrared continuum emission peak \citep{bar98}, 
high-excitation, shocked molecular hydrogen emission \citep{dav94,dav95},
and the optically visible shock-excited gas of HH 197 \citep{bal97}.

Extrapolating from the vicinity of HH 197 along 
P.A. $\sim$ 127$^{\circ}$, which is the 
axis  of the blueshifted L1448C CO outflow after its deflection near HH
197 through knots I, R, and S (Davis \& Smith 1995; Eisl\"offel 2000),
leads directly to HH 267. The measured radial
velocities of the  HH 267 knots
(HH 267A: V$_{LSR} = -50$ km s$^{-1}$;
HH 267B: V$_{LSR} = 0$ km s$^{-1}$;
HH 267C: V$_{LSR} = -63$ km s$^{-1}$ $-$ Bally et al. 1997)
and the velocity extent of the blueshifted CO observed by
Bachiller et al. (1990) agree well.
The reported
terminal velocity for the L1448C flow is 70 km s$^{-1}$ (Bachiller et al. 1990,
1995).
These two facts led to the suggestion that
HH 267 may be powered by L1448C \citep{bar98}. 

L1448N(A) and L1448N(B) form a close (7$^{\prime\prime}$ separation)
protobinary (Terebey \& Padgett 1997), which is unresolved in our OTF map.
The redshifted portion of the L1448N(A) flow was first detected via its
associated low-excitation molecular hydrogen emission \citep{dav95}.
Its exciting
source was first correctly identified by Barsony et al. (1998).
The corresponding blueshifted lobe has been detected only in
an extended, conical reflection nebulosity whose peak is estimated to lie 
$1.^{\prime\prime}5$ west and 6$^{\prime\prime}$ north of L1448N(A) (Bally
et al. 1993).
The position angle of $\sim 150^{\circ}$ for this flow
is determined by the symmetry axis of the U-shaped shocked molecular
hydrogen emission (Barsony et al. 1998; see also Eisl\"offel 2000)
and from the axis of the highly-collimated redshifted CO jet driven by
L1448N(A) as seen in the V$_{LSR}$ $=$ $+$8 km s$^{-1}$
outflow channel map of Bachiller et al. (1995). 

Redshifted gas associated with the L1448N(B) flow first appears in the
paper reporting the discovery of the L1448C outflow (Bachiller et al. 1990).
The corresponding blueshifted outflow lobe from L1448N(B) was partially
mapped by Bontemps et al. (1996).
Barsony et al. (1998) noted that HH 196, a series of blueshifted optical
emission knots 
(HH 196A: V$_{LSR} = 0$ km s$^{-1}$;
HH 196B: V$_{LSR} = 0$ km s$^{-1}$;
HH 196C: V$_{LSR} = -35$ km s$^{-1}$;
HH 196D: V$_{LSR} = -37$ km s$^{-1}$ $-$ Bally et al. 1997)
lie along the L1448N(B) outflow axis.
The position angle of the L1448N(B) outflow is 
P.A. $\sim$129$^{\circ}$.

About 20$^{\prime\prime}$ northwest of L1448N(A) lies L1448NW. Recent
observations suggest that L1448NW drives a small-scale H$_2$ outflow
along an east-west direction (Eisl\"offel 2000).
Although we
do not detect a CO outflow associated with L1448NW with our spatial
resolution, unpublished interferometric observations do 
indicate the presence of an E-W flow centered on L1448NW (Terebey 1998).

\subsubsection{The L1448 IRS 2 Core}

L1448 IRS 2 was confirmed as a Class 0 protostar by O'Linger et al. (1999),
who reported a CO outflow associated with this source. The outflow's 
symmetry axis along P.A. $\sim$ 133$^{\circ}$ and full opening angle, $\phi\
=$27$^{\circ}$,
were initially inferred from 
(1) the locations of HH 195A$-$D, (2) a fan-shaped reflection nebulosity
emanating from IRS 2 in the K$^{\prime}$ images of Hodapp (1994), (3) the
positions of CO ``bullets'', detected at the 3$\sigma$ level along the
outflow axis about 10$^{\prime}$ to the northwest of IRS 2, and (4) the 
apparent V-shaped morphology of the blueshifted CO emission.
A more recent, H$_2$ image of this region has led to a
refinement of the IRS 2 outflow axis determination to P.A. $\sim$ 138$^{\circ}$
(Eisl\"offel 2000).

\subsection{Velocity Maps}

In order to elucidate the velocity structure of the CO emission,
we present Figures 4, 5, \& 6.
All three figures are presented in rotated coordinates, such that the 
major axis of our map along P.A. $=$ 135$^{\circ}$
now lies horizontally.
Figure 4 is meant to be used as a key to identify the HH objects (crosses)
and Young Stellar Objects (YSO's--open stars for Class 0 protostars, filled
square
for the Class I protostar, L1448 IRS1) indicated by the same symbols in
the CO veolocity channel maps of Figures 5 \& 6.
Our beamsize is indicated in the lower right-hand corner of each panel.

Figure 5 shows the contoured greyscale images of the blueshifted 
integrated CO intensities for five velocity intervals blueward 
of the  ambient cloud velocity, 
proceeding top down from the highest
velocities in the top panel, to the lowest, cloud-core velocities
in the bottom panel.
Figure 6 shows the same for the redshifted emission.
In order to obtain good signal-to-noise, our highest velocity channel
maps have been integrated over a 4 km s$^{-1}$ velocity
interval. The other channel maps have been integrated over 2 km s$^{-1}$
intervals.
These maps shed more light on the emission features enumerated
in \S 3.1, and uncover
additional information that is not
obvious from the outflow map of Figure 2.
It is immediately apparent from these maps that much of the emission
within the line core delineates gas that has been entrained in the outflows.
In particular, the $\sim 15^{\prime}$ long blueshifted feature,
which extends from IRS 2 to HH 193,
is seen to some degree in all of the maps depicting emission blueward of
the ambient cloud velocity (Figures 5a$-$e),
as well as in the core gas redward of the ambient cloud velocity
(Figures 6d \& e).
Redshifted, as well as blueshifted, CO emission surrounds HH 193. 
This is not surprising, since the radial velocities
and linewidths of HH 193A, B, and C, are -18 \& 40 km s$^{-1}$, 10 \& 70
km s$^{-1}$, and -10 \& 50 km s$^{-1}$, respectively (Bally et al. 1997).
This significant overlap of blueshifted and redshifted 
emission strongly suggests that the blueshifted
feature is oriented close
to the plane of the sky.

The blueshifted feature is part of a longer feature, which is bipolar
about IRS 2 at P.A. $\sim$ 152$^{\circ}$. The blueshifted emission is visible as
a well-defined structure from V$_{LSR} = $-8 to 0 km s$^{-1}$. Similarly,
redshifted emission is visible as a well-defined structure
from V$_{LSR} = $8 to 16 km s$^{-1}$.
At the highest velocities (Figures 5a \& 6a), the
blue- and redshifted emission is highly collimated
along P.A. $\sim 152^{\circ}$, centered on L1448 IRS 2.
However, the blueshifted emission intersects
a second blueshifted feature emanating from the
L1448N/L1448C region along P.A. $\sim$ 129$^{\circ}$ (Figure 5a).
The intersection occurs $\sim$2$^{\prime}$ 
downstream of the HH 196 knots.
A third blueshifted feature branches off to the west $\sim$3$^{\prime}$ 
downstream from the HH 196 knots
(Figure 5b).
The redshifted emission along P.A. $\sim$ 152$^{\circ}$ from IRS 2 extends
$\sim 9^{\prime}$ southeast of IRS 2 (Figures 6a$-$c), where it shows a
prominent peak
in Figure 6b, as well as in the CO total integrated
intensity outflow map in Figure 2.

The blue- and redshifted peaks adjacent to L1448 IRS 2
are most prominent in the lowest-velocity outflow emission (Figures 5c \& 6c).
The blueshifted peak is spatially coincident with HH 195E.
At higher velocities (Figures 5a \& b), the blueshifted emission extends 
along a P.A. $\sim$ 125$^{\circ}$ from IRS 2, past IRS 1, and
may continue past HH 194,
which shows a local peak in blueshifted emission (Figures 5b \& c).
Curiously,
the HH 194 knots are all {\it redshifted}, with very large linewidths (HH 194A:
V$_{LSR} =$66 km s$^{-1}$, $\Delta$V$=$110 km s$^{-1}$; HH 194B:
V$_{LSR} =$66 km s$^{-1}$, $\Delta$V$=$150 km s$^{-1}$; HH 194C:
V$_{LSR} =$110 km s$^{-1}$, $\Delta$V$=$60 km s$^{-1}$).
HH 194C has been associated with redshifted outflow emission from
IRS 1 (Bally et al. 1997).
Although there is a prominent local peak in the blueshifted
CO emission at the position of HH 194,
there is no evidence of
redshifted CO emission (Figures 6a$-$c) at this position.

The extended redshifted emission directly to the southeast
of L1448N/L1448C is also clearly present at core velocities, very strongly
so in Figure 6d, and even in Figure 5e. The previously suggested structure
along three separate axes is also seen in Figures 6a$-$c; mostly strikingly,
in Figure 6b. These three axes, along P.A. $\sim$ 180$^{\circ}$,
150$^{\circ}$, \& 129$^{\circ}$,
correspond to the PAs of the redshifted lobes of
the outflows associated with L1448C, L1448N(A), \& L1448N(B), respectively.
The redshifted feature that peaks on HH 277 is most
prominent as a separate velocity feature in Figure 6c.

In the next section, we consider these features in connection with what is
known about the individual outflows in order to interpret the outflow
morphology in L1448.

\section{DISCUSSION}

\subsection{Interpretation of Outflow Structure}

Figures 7$-$9 present the various outflow extents and position angles.
Figure 7 is relevant for the discussion of alternative interpretations
of the outflow emission centered on the L1448 IRS 2 ammonia core.
Figure 8 is used for the discussion of the outflows originating from
sources embedded in the ammonia core associated with L1448C and L1448N.
For both Figures 7 and 8, the outflow axes and extents
are superposed on our integrated high-velocity CO linewing map
of the L1448 cloud. Figure 9 shows the same outflow
axes superposed on a much higher spatial-resolution ($\sim$ 1$^{\prime\prime}$
vs. $\sim$ 55$^{\prime\prime}$) H$_2$ image of the L1448 region from Eisl\"offel
(2000).
(This is the only figure using J2000 coordinates.)
In all cases (except for the two newly-identified outflow features seen in
our CO maps associated with L1448 IRS 2 and the unidentified source
outside our map boundaries), outflow position angles were determined
from previously-published, arcsecond-scale outflow data.
For all of the outflows, we find 
good agreement between large-scale
CO features seen in our maps and flow axes that have been determined from the
previous, higher-resolution observations.
In  Figures 7$-$9, {\it solid}, colored
lines denote well-established flow position angles and extents,
derived from our own CO data and the published literature, whereas
{\it dashed} lines denote outflow position angles and extents that are
consistent with our new CO data.

\subsubsection{Outflow Emission from the L1448 IRS 2 Ammonia Core}

High-velocity CO outflow activity centered on L1448 IRS 2 was discovered
from low-spatial resolution ($\approx$ 55$^{\prime\prime}$) mapping (O'Linger et
al. 1999).
These authors suggested that L1448 IRS 2 was the source
of a single outflow, with a constant 
opening angle, as depicted in Figure 7a. The previous outflow symmetry
axis along P.A. $\sim$ 133$^{\circ}$ and opening angle, $\phi \sim 27^{\circ}$,
were derived from the high spatial resolution K$^{\prime}$ images of Hodapp
(1994).
We derive a new outflow symmetry axis along P.A. $\sim$ 138$^{\circ}$ from 
the more recent H$_2$ images of Eisl\"offel (2000). 
Therefore, we have been able to determine more accurately the position angles
for the proposed outflow cavity walls, which should lie along P.A. $\sim$ 152$^{\circ}$ and 
P.A. $\sim$ 125$^{\circ}$, if the IRS 2 outflow retains its
initial opening angle out to large distances.
This model explains the presence of high-velocity blue- and redshifted CO
emission seen along these position angles at large distances from IRS 2, 
notably the V-shaped
morphology of the blueshifted gas and the presence of several CO ``bullets''
located along the proposed outflow axis, well beyond HH 195A$-$D.
In a constant opening angle outflow scenario,
HH 193 lies along one arm of this V, at P.A. $\sim$ 152$^{\circ}$.
The blueshifted gas along the other arm of the V (P.A. $\sim$ 125$^{\circ}$)
would be confused with emission from the E-W outflow in this vicinity,
but since the blueshifted emission extends
{\it past} IRS 1 towards the {\it redshifted} HH 194 knots, at least part of
this emission could be due to IRS 2. 
Although there is redshifted CO emission along P.A. $\sim$ 138$^{\circ}$ and
along P.A. $\sim$ 152$^{\circ}$, there is little evidence for redshifted
emission
along P.A. $\sim$ 125$^{\circ}$. However, this
may be due to confusion with the three redshifted lobes associated with
L1448C, L1448N(A), \& L1448N(B).

Figure 9 clearly shows an outflow associated
with IRS 2 along a P.A. $\sim$ 138$^{\circ}$.
The redshifted gas along the outflow axis is prominent
in the H$_2$ emission, but is not clearly apparent in our CO maps beyond
the redshifted peak about 1$^{\prime}$ southeast of IRS 2. Hints of more
extended emission along P.A. $\sim$ 138$^{\circ}$ may be seen, 
however, in Figures 6b \& c, and
curiously, in {\it blueshifted} emission extended along this axis to the
southeast of IRS 2 in Figure 5d. Such overlap of blueshifted gas along the
redshifted outflow axis is expected for outflows oriented nearly in the
plane of the sky.
Indications of extended blueshifted emission along P.A. $\sim$ 138$^{\circ}$,
at least a few arcminutes
downstream of HH 195A$-$D, are seen in Figures 5b \& c.

However, there are a few problems with the single outflow, constant opening
angle model:
(1) Figure 9 indicates that although
the initial opening angle of the IRS 2 outflow is
$\phi \sim 27^{\circ}$, the two strands of H$_2$ defining this opening angle
join in a bow shock structure in the vicinity of HH 195A-D;
(2) The CO data show little evidence for emission along the cavity wall
at P.A. $\sim$ 125$^{\circ}$, although we note that there is much confusion
from high-velocity gas associated with other outflows along this position
angle to both sides of IRS 2;
(3) HH 193 lies precisely at the end of the outflow wall in this model,
an unlikely location for a shock;
(4) The velocity dispersion of the blueshifted feature along
P.A. $\sim$ 152$^{\circ}$ is high, since it is prominent at both ambient
cloud velocities and at highly blueshifted velocities (Figures 5, 6d\&e);
(5) The highest-velocity outflow emission should converge
towards the outflow axis (P.A. $\sim$ 138$^{\circ}$) due to projection effects,
but the highest-velocity outflow emission lies along P.A. $\sim$ 152$^{\circ}$,
the proposed outflow wall.

Our maps are not sensitive enough to have picked up the
highest-velocity outflow emission, however, which is severely diluted
in our large ($\sim$1$^{\prime}$) beam. Nevertheless, the absence of CO emission
along P.A. $\sim$ 138$^{\circ}$ in the highest velocities suggests that
the feature along P.A. $\sim$ 152$^{\circ}$ defines a separate outflow axis.
This has led to an alternate interpretation of the high-velocity CO
associated with L1448 IRS 2 in which the presence
of {\it two} outflows is required, as depicted in Figure 7b,
with one outflow along P.A. $\sim$ 138$^{\circ}$,
and a new, second
outflow along P.A. $\sim$ 152$^{\circ}$, so prominent in the CO data.
In this scenario, the new IRS 2 outflow would be responsible for exciting
HH 193. Two outflows,
along distinctly different position angles, would also suggest that IRS 2
is a binary system.
Although there is currently no evidence to indicate this source
to be binary from available continuum data obtained with the 
Submillimetre Common User Bolometer Array (SCUBA) at the JCMT on Mauna Kea,
Hawaii
(O'Linger et al. 1999), it is
possible that IRS 2 is a
compact binary on a scale smaller than 7$^{\prime\prime}$ (the resolution
of the SCUBA 450 $\mu$m data). Recent work indicates a high
incidence of binarity among young stellar systems (eg., Ghez, Neugebauer,
\& Matthews 1993;
Looney, Mundy, \& Welch 2000).
Only arcsec/sub-arcsec
imaging at either millimeter or centimeter wavelengths could test the
binary hypothesis further. 

Evidence of other outflow activity near L1448 IRS 2 
is found by noting that the very confined outflow, 
whose lobes peak only $\sim$1$^{\prime}$ on either side of IRS 2, has its
redshifted peak well-aligned along the P.A. $=$ 138$^{\circ}$
outflow that excites HH 195A-D, but the corresponding 
blue peak closest to IRS 2 is skewed at a somewhat shallower position
angle closer to P.A. $\sim$ 125$^{\circ}$.
No CO emission (Figure 2, Figures 6a$-$c)
is seen along P.A. $\sim$ 125$^{\circ}$ on the opposite side from IRS 2,
which would
be expected if there were an outflow along this direction.
The blue peak
is spatially coincident with HH 195E, however, the only HH 195 knot which is off
the
P.A. $=$ 138$^{\circ}$ axis of the IRS 2 outflow.
It has been argued that IRS 1 drives an east-west oriented
outflow and is the most probable driving source of HH 195E (Bally et al. 1997;
Eisl\"offel 2000).
Although the presence of such an E-W oriented outflow in this region is
undisputed,
it is possible that an as yet undiscovered source, other than IRS 1, may
be the responsible agent.
Thus, the positioning of the blue peak so close to IRS 2
may be coincidental, and due primarily to local heating associated with
HH 195 E, and overlapping outflows associated with
blueshifted emission from
IRS 2 and IRS 1. This picture is supported by the L1448 H$_2$ mosaic of
Eisl\"offel (2000),  shown in Figure 9, which shows the H$_2$
emission in the vicinity of HH 195E pointing toward IRS 1, not IRS 2.

\subsubsection{Outflow Emission from the L1448C/L1448N Ammonia Core }

The position angles of the blue- and red-shifted outflowing gas
powered by L1448C are indicated by the green lines in both Figures 8 \& 9.
Solid green lines indicate the L1448C outflow's direction and extent
as determined by previous workers (see the caption of Figure 8 for references).
The dashed green line on the blue-shifted side
represents the continuation of the L1448C outflow proposed
by Barsony et al. (1998). 
The blueshifted L1448C
outflow suffers a large deflection from its original direction, as
can be seen clearly in Figure 9, which shows
the blueshifted L1448C outflow axis passing directly through
H$_2$ emission knots I \& S, and about 20$^{\prime\prime}$ north of emission
knot R.
Emission knot R lies within more extended H$_2$ emission
that appears to form a U-shaped or bow-shock structure which opens toward the
southeast, bisected by the L1448C outflow axis. The sides of the U are separated
by $\sim$ 40$^{\prime\prime}$,
with the brighter side (including knot R) lying to the south of the
L1448C outflow axis.

This final, deflected, blue-shifted  outflow axis lies along
P.A. $\sim$ 127$^{\circ}$, where 
a long finger of high-velocity blueshifted CO emission is found. 
Blueshifted CO emission surrounds the HH 267 complex in a horseshoe shape
about the proposed extension of the L1448C outflow, suggesting the outflow
may be responsible for this emission. The dashed green line on the redshifted 
side in Figure 8 indicates the possible continuation of the L1448C outflow
to the south.

The L1448N(A) molecular outflow axis and extent are depicted by the 
purple line in Figures 8 \& 9.
Only redshifted molecular gas associated with the L1448N(A) outflow,
along P.A. $\sim$ 150$^{\circ}$,
is  detected in our CO maps, with a total length
of at least 0.7 pc, as seen in Figure 8.
The U-shaped molecular hydrogen emission that traces part of the redshifted
outflow cavity wall from the L1448N(A) outflow, as seen in Figure 9,
is unresolved in our CO maps. Knots of H$_2$ emission which trace the outflow
wall are 
$\le$30$^{\prime\prime}$ apart as far as 3$^{\prime}$ to the southeast of
L1448N(A) (Barsony et al. 1998; Eisl\"offel 2000).
Taken together with the observed length of the CO redshifted lobe,
this suggests a
collimation factor of at least 16:1.
The opening half-angle of the L1448N(A) outflow was estimated
to be $\phi/2\ \approx\ 25^{\circ}$,  
from the morphology of the near-infrared reflection nebulosity
to the north of L1448N(A),
associated with what would be the blueshifted outflow lobe (Bally et al. 1993).
The blueshifted flow powered by L1448N(A) is not apparent in our CO maps,
judging by the drop in the blueshifted, high-velocity CO contour levels along
the symmetry
axis of its NIR reflection nebula. The lack of blueshifted CO emission from the
L1448N(A)
outflow is most likely accounted for by the likelihood
that the cloud boundary has been reached in this direction, and that the flow
has broken out of the molecular cloud. 

The L1448N(B) molecular outflow axis and extent are depicted by the 
mustard-colored lines in Figures 8 \& 9.
The position angle, P.A. $\sim$ 129$^{\circ}$, of the L1448N(B) outflow
was determined by the orientation of the redshifted CO outflow
driven by L1448N(B) from Figure 1b and noting that this CO flow symmetry axis
intersects HH 196 (Barsony et al. 1998).
The true spatial extent of the L1448N(B) CO outflow, however, 
is demonstrated here for the first time. On the scale of our map, 
the L1448N(B) outflow remains unresolved along its width.
The optical emission knots of HH 196
are, indeed, found to lie right along the L1448N(B) CO outflow axis,
confirming the identification of L1448N(B) as their exciting source
(Figure 8).
Approximately
2$^{\prime}$ downstream of HH 196, the L1448N(B) flow becomes confused
in projection with the P.A. $\sim$ 152$^{\circ}$ outflow from IRS 2.
If the outflow continues along P.A. $\sim$ 129$^{\circ}$,
it could account for the bulges in the
blueshifted emission
to the south and southwest of HH 193 (Figure 2, Figure 5c, Figure 8), and might
even be responsible for the ``C''--shaped blueshifted emission structure
east of the HH 267 system, $\sim$2.5 parsecs from L1448N(B).
The redshifted lobe associated with this source appears to terminate 
$\sim$2$^{\prime}$ northwest of HH 277, although this could be due to 
confusion with the high-velocity CO emission surrounding HH 277,
which seems to be part of an outflow driven by an unidentified source
outside of the boundaries of our map.
Using the most conservative length for the L1448N(B) outflow, the major axis
taken from the HH 196 knots through the end of the redshifted lobe
($\sim$12$^{\prime}$), the derived lower limit for the collimation factor
is $\ge$12:1. 
Estimates of the initial opening angle and width of the
L1448N(B) outflow await higher spatial resolution, interferometric imaging.

A dark blue dashed line in Figure 8 indicates a possible outflow axis passing
through an otherwise unexplained, high-velocity redshifted feature associated 
with HH 277, in the southeast quadrant of our CO map.
This redshifted CO velocity feature is most prominently seen in Figure 6c. The 
orientation of this structure is almost perpendicular to the general orientation
of our map. Thus, HH 277 is probably driven by a source off the edge of our map.

Finally, although the 
origin of the HH 267 knots cannot definitively
be resolved based on the CO data we present here, our observations
do constrain the driving source.
L1448N(A) \& (B) can be ruled out as possible driving sources of HH 267,
since the P.A.'s of their associated outflows are along completely different
directions
than the lines linking them with HH 267.  Furthermore, blue-shifted molecular
gas has yet to be detected from L1448N(A).
Terminal velocities for the L1448 IRS 2 outflows can not be determined from
our data, but
the P.A.'s of both of these outflows also miss 
the HH 267 complex completely.
However, the P.A. of
the deflected blueshifted lobe of the L1448C flow
goes right through HH 267, and the reported terminal velocity for this outflow
(70 km s$^{-1}$: Bachiller et al. 1990, 1995) is in good agreement with the
measured HH 267 velocities (Bally et al. 1997), as suggested by Barsony et al.
(1998).

\subsection{Cloud Dispersal by Giant Protostellar Flows?}

The most dramatic evidence for the direct effects of
the outflows on the L1448 molecular cloud is seen in the
distortions of the cloud contours at all velocities in Figures 5 \& 6.
Although we cannot estimate the masses and energetics of each
individual outflow in our maps due to confusion in space and velocity,
we can, nevertheless,
estimate the {\it total}
contribution of the outflows to the cloud's energetics.
The Local Thermodynamic
Equilibrium (LTE) analysis used to estimate the combined mass of the outflows
(M$_{tot}\approx 0.7$ M$_{\odot}$) is discussed in O'Linger et al. (1999).
Optically thin high-velocity CO emission was assumed, given the lack of
observed high-velocity $^{13}$CO emission in the velocity intervals outside
the cloud core velocities (see Figure 3).
Therefore,
the resultant derived mass is a strict lower
limit, since no attempt was made to correct for
the considerable mass expected to be masked by the line core emission.

For highly inclined outflows ($i>70^{\circ}$),
the characteristic velocity which is used
to calculate outflow energetic parameters is best chosen as
the geometrical mean between the highest observed velocities, $V_{CO}$, and the
inclination-corrected velocity, $V_{CO}/cos(i-{{\phi}\over {2}})$,
where ${\phi}\over {2}$ is the half-opening angle of the outflow
(Cabrit \& Bertout 1992).
Assuming the outflow inclinations, 70$^{\circ}$ $\le$ $i$ $\le$ 90$^{\circ}$,
V$_{char} = 22 - 34$
km s$^{-1}$ for the L1448 outflows,
the total momentum in all the flows 
is computed to be
16 M$_{\odot}$ km s$^{-1}$ $\le$ M$_{tot}$V$_{char}$ $\le$
24 M$_{\odot}$ km s$^{-1}$. This range of values is nearly equivalent to
the momentum content of the quiescent NH$_3$ cores, assuming 50 M$_{\odot}$
total cores and $v_{turb}$ $\sim$ 0.5 km s$^{-1}$ \citep{ba86a}.
Even more striking, the total kinetic energy in all the
flows,
2$\times 10^{45}$ ergs $\le$ $1\over {2}$M$_{tot}$V$_{char}^2$ $\le$
8$\times 10^{45}$ ergs,
exceeds the gravitational binding energy ($\sim\ GM^2/R\ \approx$ 5 $\times$
10$^{44}$ ergs)
of the NH$_3$ cores by an order of magnitude, and 
the gravitational binding energy of the 100 M$_{\odot}$ C$^{18}$O cloud
(9 $\times$ 10$^{44}$ ergs),
contained within a 1.3 pc $\times$ 0.7 pc region \citep{ba86b},
by a factor of five.

The total outflow momentum quoted above is, in fact, a 
lower limit, since these outflows are still gaining momentum from the force
provided by the central driving engines, and the total outflow mass 
may be grossly underestimated.
The magnitude of both the total energy and momenta of the outflows
suggests these outflows are capable of dispersing the NH$_3$ cores,
with the caveat that it is unclear, both from the outflow and ambient cloud
morphology, how the outflow momenta can be adequately transferred to
the surrounding core. Possibly, this can be accomplished as the individual
outflow opening angles increase with time.

\section{Summary}

$\bullet$ Spectral-line ``on-the-fly'' mapping was used at
the NRAO 12-meter millimeter telescope to produce a large-scale 
(47$^{\prime}$ $\times$ 7$^{\prime}$) CO (J$=$1$\rightarrow$0)
map of the L1448 dark cloud,
sensitive enough (1 $\sigma$ $=$ 0.1K) to
enable the detection of outflow activity on parsec-scales and the 
identification of six distinct molecular outflows.
Large-scale, high-spatial resolution optical and near-infrared
images of shocked gas emission regions associated with each outflow were
crucial for identifying the CO counterparts of these outflows.
Three of the outflows are associated with
the Class 0 protostars, L1448C, L1448N(A), \& L1448N(B). Two outflows
are associated with the Class 0 protostar, L1448 IRS 2.
A sixth outflow, apparently associated
with HH 277, is probably driven by an unidentified 
source located outside of our map.

$\bullet$
For all of the outflows which have previously been identified through
high-resolution interferometric observations or small-scale shocked gas
emission, we find
good agreement between large-scale
CO features and flow axes that have been determined from these
higher-resolution
observations.

$\bullet$ We find evidence
of two distinct outflows emanating from the recently-confirmed Class 0
protostar, L1448 IRS 2 (O'Linger et al. 1999),
suggesting that IRS 2 is an unresolved binary system.
One of these outflows lies along P.A. $\sim$ 138$^{\circ}$, and is apparent both
in H$_2$ emission (Eisl\"offel 2000) and, to a lesser degree, in our CO data.
The second outflow lies along P.A. $\sim$ 152$^{\circ}$ and is seen as a
highly-collimated jet in our CO data, culminating, on the blueshifted side,
in HH 193.
 
$\bullet$ The ambient cloud emission contours are severely disturbed by the 
outflows, suggesting a large fraction of the ambient cloud in the mapped
region has been entrained in, or stirred up by, the outflows.

$\bullet$ The 
total outflow kinetic energy
($>$ 2 $\times$ 10$^{45}$ ergs)
and combined outflow momenta
($>$ 16 M$_{\odot}$ km s$^{-1}$)
indicate that the outflows are energetically, and probably dynamically, capable
of dispersing the dense ammonia cores out of which the 
protostellar outflow driving sources of five of the six identified
flows,  L1448C, L1448N(A), L1448N(B),
and L1448 IRS 2, are currently forming.
However, it is unclear,
from both the outflow and ambient cloud morphology,
how the outflow momenta can be adequately transferred to the surrounding 
cores.

ACKNOWLEDGEMENTS:  We thank Dr. Darrel Emerson, Dr. Eric Greisen, and Dr. Jeff
Mangum of NRAO for the
development, implementation,
and improvement of the spectral-line On-The-Fly mapping capability of the
12-meter telescope. 
GWC, JO, and MB gratefully acknowledge financial support from NSF grant
AST-0096087 while part of this work was carried out. 
Part of this work was performed while GWC
held a President's Fellowship from the University of California.  MB's NSF POWRE
Visiting Professorship at Harvey Mudd College, NSF AST-9731797, provided the
necessary time to bring this work to completion.
JO acknowledges financial support by the NASA Grant to the Wide-Field 
Infrared Explorer Project at the Jet Propulsion Laboratory, California
Institute of Technology.  We would like to thank our referee, John Bally,
for his many helpful suggestions which greatly improved this paper.

\newpage
%
%
\figcaption[Wolf-Chase.fig1.ps]{{\bf High-velocity CO Outflow Emission on
Large and Small Scales in the L1448 Dark Cloud}} 
{\bf a.} Blue contours indicate high-velocity blueshifted
($-12.1\ \le\ V_{LSR}\ \le\ -1$ km s$^{-1}$) emission and 
red contours indicate high-velocity redshifted
($+$8.1 $\le\ V_{LSR}\ \le\ $+$17.8$ km s$^{-1}$)
CO J$=$1$\rightarrow$0 emission 
in the 47$^{\prime}\ \times\ 7^{\prime}$ region
we mapped with the NRAO 12-meter telescope.
Contour levels start at 2 K km s$^{-1}$ ($\approx$ 3 $\sigma$), and increase
in 1.5 K km s$^{-1}$ intervals. 
For comparison, the dashed rectangle shows the approximate area
previously mapped (in CO J$=$2$\rightarrow$1), shown in the inset.
Stars indicate the positions of L1448C, L1448N, L1448NW, and L1448 IRS 2. 
All of these are Class 0 sources; L1448N is a 7$^{\prime\prime}$
separation protobinary, consisting of L1448N(A) and L1448N(B). L1448NW
lies $\sim$ 20$^{\prime\prime}$ to the northwest of the protobinary.
The filled box indicates the position of the Class I source, L1448 IRS 1.
The 55$^{\prime\prime}$ FWHM NRAO beamsize
is indicated in the lower right-hand corner.
{\bf b.} The previous CO J$=$2$\rightarrow$1 IRAM 30-meter
map of high-velocity gas in L1448 (from \citep{bac90}):
 Solid contours indicate high-velocity,
blueshifted ($-$ 55 km s$^{-1}$ $\le$  V$_{LSR}$ $\le$ 0 km s$^{-1}$) gas,
and dotted contours indicate high-velocity, redshifted
($+$ 10 km s$^{-1}$ $\le$  V$_{LSR}$ $\le$ $+$ 65 km s$^{-1}$) gas. 
First contour and contour intervals are at 10 K km s$^{-1}$.
In addition to the famous protostellar outflow powered by L1448C,
the weaker outflow, powered by L1448N(B), is also detected . 
The 12$^{\prime\prime}$ FWHM IRAM beamsize is indicated in the lower
right-hand corner.
%
%
%
\figcaption[Wolf-Chase.fig2.ps]{\bf CO Outflow Integrated Intensity Map} 
Names and positions of all the Herbig-Haro objects (black crosses) are shown, 
as well as the five Class 0 sources (black stars) and the Class I source, 
L1448 IRS 1 (solid black box). Velocity intervals and contour levels are the
same as in Figure 1a.

%
%
\figcaption[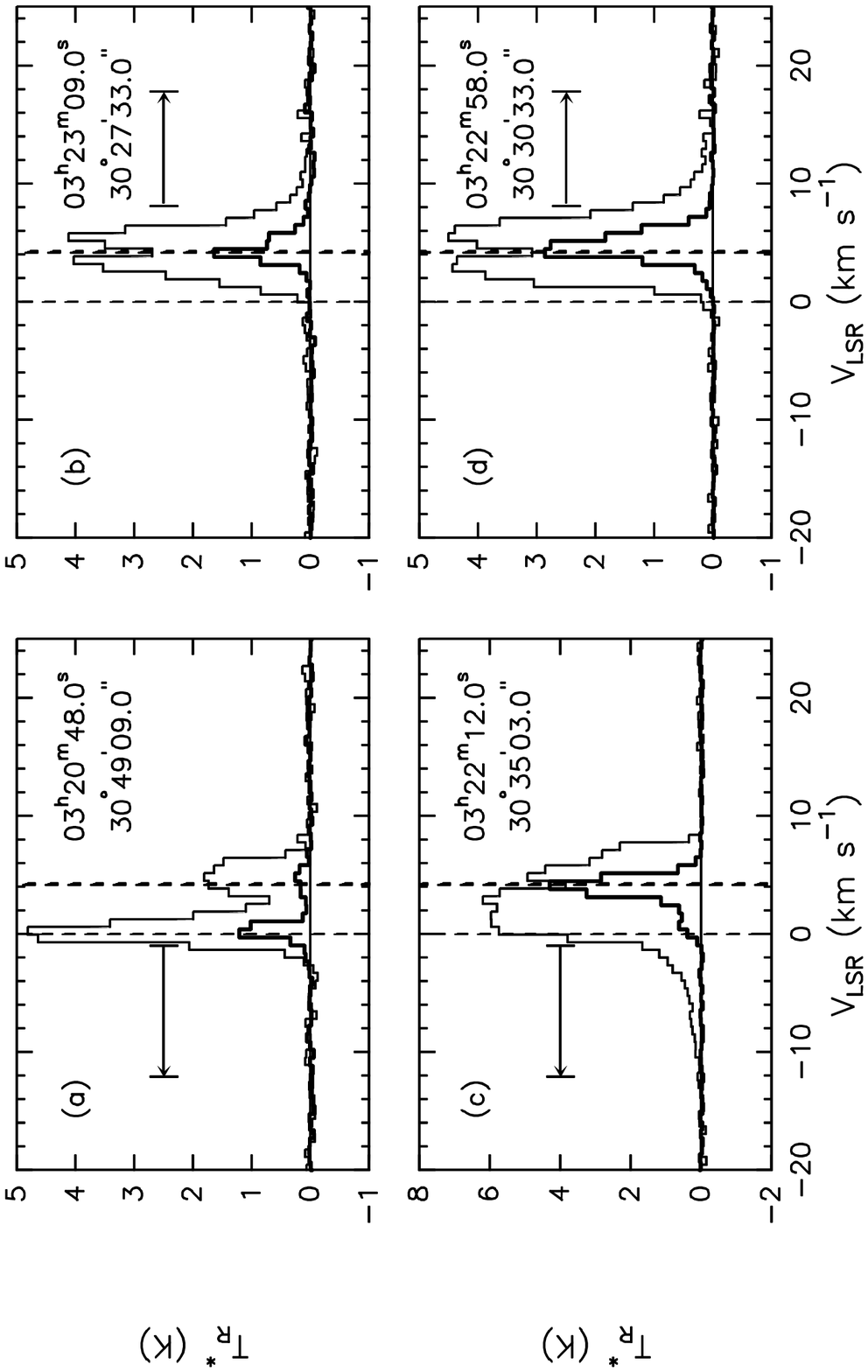]{\bf Isotopic CO Spectra} 
$^{12}$CO (thin solid line)
\& $^{13}$CO (thick solid line) J=1$\rightarrow$0 spectra
obtained at the four positions on our map whose B1950 coordinates are indicated:
(a) CO spectra near the HH 267 knots 
(just off the northwest corner of our map)
clearly show two separate velocity components:
a brighter component at the same velocity as HH 267B,
V$_{LSR}=$0 km s$^{-1}$ (thin vertical dashed line),
and a dimmer component
at V$_{LSR}=$4.25 km s$^{-1}$ (thick vertical dashed line).
(b) $-$ (d) CO spectra show
strong CO self-absorption at the ambient
cloud velocity, V$_{LSR}=$4.25 km s$^{-1}$.
Arrows indicate the velocity ranges used
to determine integrated intensities, masses, and energetics for
the outflow emission.
%
%
\figcaption [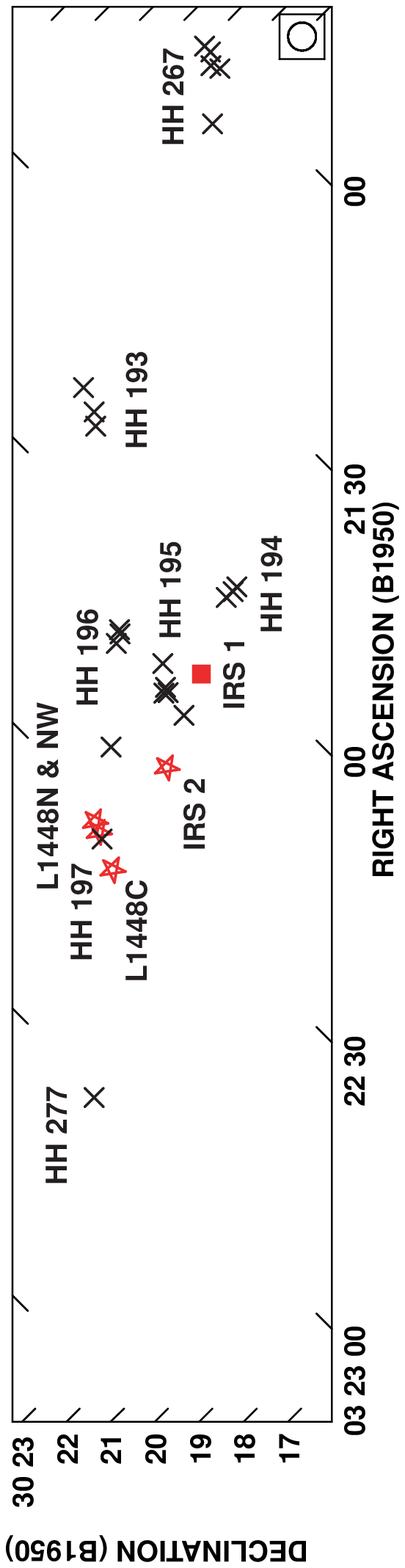]{\bf Template for Figures 5 \& 6} 
The positions
of all the Herbig-Haro objects are indicated by tilted black crosses;
the Class I source, L1448 IRS 1, is shown by a filled red box; the Class 0
protostars, L1448 IRS 2, L1448C, L1448N(A) \& (B), and L1448NW are indicated
by red stars. The separation between L1448C \& L1448N is $\sim$ 
80$^{\prime\prime}$.
%
%
\figcaption [Wolf-Chase.fig5.ps]{\bf Velocity Channel Maps of Blueshifted Gas}
Contoured greyscale images of the 
integrated CO intensity over velocity intervals blueward of
the ambient cloud emission. Velocity intervals, lowest contour levels,
and contour intervals, are, respectively:
(a) -8 to -4 km s$^{-1}$, 1 K km s$^{-1}$, 0.5 K km s$^{-1}$;
(b) -4 to -2 km s$^{-1}$, 1 K km s$^{-1}$, 0.5 K km s$^{-1}$;
(c) -2 to 0 km s$^{-1}$, 1.5 K km s$^{-1}$, 1 K km s$^{-1}$;
(d) 0 to 2 km s$^{-1}$, 2 K km s$^{-1}$, 2 K km s$^{-1}$;
and (e) 2 to 4 km s$^{-1}$, 2 K km s$^{-1}$, 2 K km s$^{-1}$.
Positions of Herbig-Haro objects (tilted yellow crosses), IRS 1 (filled red
box), and Class 0 objects (red stars) are indicated.
%
%
\figcaption [Wolf-Chase.fig6.ps]{\bf Velocity Channel Maps of Redshifted Gas}
Contoured greyscale images of the 
integrated CO intensity over velocity intervals redward of
the ambient cloud emission. Velocity intervals, lowest contour levels,
and contour intervals, are, respectively:
(a) 12 to 16 km s$^{-1}$, 1 K km s$^{-1}$, 1 K km s$^{-1}$;
(b) 10 to 12 km s$^{-1}$, 1 K km s$^{-1}$, 1 K km s$^{-1}$;
(c) 8 to 10 km s$^{-1}$, 2 K km s$^{-1}$, 1 K km s$^{-1}$;
(d) 6 to 8 km s$^{-1}$, 2 K km s$^{-1}$, 2 K km s$^{-1}$;
and (e) 4 to 6 km s$^{-1}$, 2 K km s$^{-1}$, 2 K km s$^{-1}$.
Positions of Herbig-Haro objects (tilted yellow crosses), IRS 1 (filled red
box), and Class 0 objects (red stars) are indicated.
%
%
\figcaption [Wolf-Chase.fig7.ps]{\bf Alternate Models
for the CO Outflow Emission from L1448 IRS
2}
{\bf a.} One outflow constant-opening angle model (O'Linger et al.~1999).
The outflow axis (P.A. $\sim$ 138$^{\circ}$) and corresponding extent of the
H$_2$
emission are denoted by the solid black line. The dashed black
line indicates possible extension of this outflow and is based on observed
CO bullets along this axis.
The dotted lines indicate the outflow walls, determined from a fan-shaped
reflection nebulosity ($\phi \sim 27^{\circ}$) seen in K$^{\prime}$ emission
(Hodapp 1994), as well as two strands of H$_2$ emission originating from
IRS 2 which lie along these axes (Eisl\"offel 2000). 
{\bf b.} Two outflows model. Solid and dashed black lines as in Figure 6a.
The dashed orange line indicates the axis of the 
highly-collimated bipolar CO emission which is evident in our CO data.
Positions
of all the Herbig-Haro objects (black crosses) are shown, as well as the five
Class 0 sources (black stars)
and the Class I source, IRS 1 (solid black box).
%
%
\figcaption [Wolf-Chase.fig8.ps]{\bf Outflow Axes
and Extents Superimposed on CO Map} 
Individual outflow position angles and extents superimposed
on the CO outflow integrated intensity map.
Positions of all the Herbig-Haro objects (black crosses) are shown, as well as
the five
Class 0 sources (black stars)
and the Class I source, IRS 1 (solid black box).
Position angles are indicated for the outflows associated with
L1448C (green $-$ Bachiller et al. 1995; Davis \& Smith 1995; Dutrey et al.
1997; Eisl\"offel 2000),
L1448N(B) (mustard $-$ Bachiller et al. 1990; Bontemps et al. 1996;
Barsony et al. 1998),
L1448N(A) (purple $-$ Davis \& Smith 1995; Barsony et al. 1998),
L1448 IRS 2 (black $-$ O'Linger et al. 1999;
orange $-$ this work), and high velocity gas of unknown origin
associated with HH 277 (dark blue).
The extents of the outflow lobes that are well-established from the literature
and our data are indicated with solid lines.
Features that are seen only in our CO data, and extrapolations that are
consistent with
our data, are shown with dashed lines.
%
%
\figcaption [Wolf-Chase.fig9.ps]{\bf Outflow Axes
and Extents Superimposed on H$_2$ Map} 
Individual outflow position angles superimposed
on an H$_2$ mosaic of L1448 (Eisl\"offel 2000).
Note that the axes of this figure are in J2000 coordinates.
The positions of Herbig-Haro objects (orange crosses), Class 0 sources
(red stars), and IRS 1 (open red box), are indicated.
Position angles of outflows are indicated as in Figure 8. Note that after its
bend at emission knot I, the position angle of the L1448C outflow passes
directly through emission knot S. Although emission knot R lies about
20$^{\prime\prime}$ south of the L1448C outflow axis, this knot appears to be
part of
extended emission that forms a U-shaped structure which opens toward the
southeast, bisected by the outflow axis. The sides of the U are separated by
$\sim$ 40$^{\prime\prime}$, with the brighter side (including knot R)
lying to the south of the L1448C outflow axis.

\newpage

\end{document}